\documentclass[aps,prd,preprint,showpacs,showkeys,superscriptaddress]{revtex4}

\usepackage{latexsym,amsmath}
\usepackage[dvips]{graphicx}

\begin{document}


\title{Chiral black hole in three-dimensional gravitational Chern-Simons}

\author{Wontae Kim}
\email[]{wtkim@sogang.ac.kr}
\affiliation{Department of Physics, Sogang University, Seoul 121-742, Korea}
\affiliation{Center for Quantum Spacetime, Sogang University, Seoul 121-742, Korea}

\author{Edwin J. Son}
\email[]{eddy@sogang.ac.kr}
\affiliation{Department of Physics, Sogang University, Seoul 121-742, Korea}

\date{\today}

\begin{abstract}
  A chiral black hole can be defined from the three-dimensional pure
  gravitational Chern-Simons action as an independent gravitational
  theory. The third order derivative of the Cotton tensor gives a
  dimensional constant which plays a role of the cosmological
  constant.  The handedness of angular momentum depends on the
  signature of the Chern-Simons coefficient. Even in the massless
  black hole which corresponds to the static black hole, it has a
  nonvanishing angular momentum.  We also study statistical entropy
  and thermodynamic stability.
\end{abstract}



\maketitle

It has been shown by Deser, Jackiw, and Templeton that the
three-dimensional Einstein-Hilbert action together with the
gravitational Chern-Simons(GCS) term called the topologically massive
gravity has nontrivial excitations even in the lower dimensional
gravity~\cite{djt}. On the other hand, Ba\~nados, Teitelboim, and
Zanelli (BTZ) have shown that the Einstein-Hilbert action with a
negative cosmological constant admits a black hole solution which has
a well-defined vacuum~\cite{btz}.  Recently, the fully combined theory
of cosmological topologically massive gravity (CTMG) has been
intensively studied in Refs.~\cite{lss,ss} to investigate
(negative energy) graviton modes. It has been shown that
the bulk modes do not vanish even at the critical
value \cite{cdww,gj,gkp}.

In this paper, we would like to find a black hole solution from the
pure GCS action as an independent gravitational theory, which is in
fact the well-known BTZ solution; however, the role of conserved
quantities should be reversed in the sense that the mass and the
angular momentum in the BTZ black hole correspond to the angular
momentum and the mass, respectively. Moreover, the third order
derivative in the GCS action gives integration constants, one of them
is a cosmological constant and the others are two conserved charges.
Then, we will calculate the statistical entropy and discuss its
thermodynamic stability.
 
The GCS action with a coupling $1/\mu$ is
\begin{equation}
S_{GCS} = \frac{1}{4\kappa^2\mu} \int d^3x \sqrt{-g} \epsilon^{\lambda\mu\nu} \Gamma_{\lambda\sigma}^{\rho} \left[ \partial_{\mu} \Gamma_{\nu\rho}^{\sigma} + \frac{2}{3}\Gamma_{\mu\tau}^{\sigma}\Gamma_{\nu\rho}^{\tau} \right], \label{action}
\end{equation}
where $\kappa^2 = 8\pi G_{3}$ is a three-dimensional Newton constant,
and $\epsilon^{\lambda\mu\nu}$is a three-dimensional anti-symmetric
tensor defined by $\epsilon^{012}=+1/\sqrt{-g}$.  Now, varying the GCS
action~(\ref{action}), an equation of motion is given by the vanishing
Cotton tensor $C^\mu_{~\nu}$,
\begin{equation}
\label{eom}
C^\mu_{~\nu} \equiv \epsilon^{\mu\alpha\beta} \nabla_\alpha \left[ R_{\nu\beta} - \frac14 g_{\nu\beta} R \right] = 0.
\end{equation}
It admits various solutions with a nontrivial curvature scalar; for
instance, a solution lifted up from a two-dimensional kink
solution~\cite{gijp}, $ds^2 = - \mathrm{sech}^4 (\sqrt{c}x/2) dt^2 +
dx^2 + \left[ dy + \mathrm{sech}^2 (\sqrt{c}x/2) dt \right]^2$, whose
curvature scalar depends on the coordinate,
$R=-3c/2+(5c/2)\,\mathrm{sech}^2 (\sqrt{c}x/2)$.

Here, we are going to focus on a solution satisfying the constraint of
$R_{\nu\beta} = (1/3) g_{\nu\beta} R$.  Then, the equation of
motion~(\ref{eom}) can be reduced to $(1/12)\epsilon^{\mu\alpha\beta}
g_{\nu\beta} \partial_\alpha R = 0$. As a particular solution, we get
$R = - 6\ell^{-2}$ with an integration constant $\ell$. Then, we can
write the equation of motion in the familiar form of $R_{\mu\nu} -
(1/2) g_{\mu\nu} R - \ell^{-2} g_{\mu\nu} = 0$.  The corresponding
action to this equation of motion is the Einstein-Hilbert action with
a cosmological constant so that its exact solution can be the BTZ
black hole~\cite{btz}.  So, the line element is naturally written as
\begin{align}
& ds^2 = -N^2(r) dt^2 + \frac{dr^2}{N^2(r)} + r^2 (d\theta - \Omega_0(r)dt)^2, \label{met} \\
& N^2(r) = -m + \frac{r^2}{\ell^2} + \frac{j^2}{4r^2} = 
\frac{(r^2-r_+^2)(r^2-r_-^2)}{r^2\ell^2}, \\
& \Omega_0(r) = \frac{j}{2r^2}=\frac{r_+r_-}{r^2 \ell},
\end{align}
where the inner and outer horizon radii are given by
\begin{equation}
r_\pm = \frac{\ell}{2} \left[ \sqrt{m+\frac{j}{\ell}} \pm \sqrt{m-\frac{j}{\ell}} \right],
\end{equation}
and the two parameters $m$ and $j$ were written as $m =
\frac{r_+^2+r_-^2}{\ell^2}, \quad j = \frac{2r_+r_-}{\ell}$.  Then,
the mass $M$ and angular momentum $J$ in the GCS action are given
by~\cite{dt,kl}
\begin{equation}
M = \frac{j}{\mu\ell^2}, \quad J = \frac{m}{\mu}. \label{ADM}
\end{equation}
Rewriting the metric functions and the two horizons $r_\pm$ in terms
of $M$ and $J$, we get
\begin{align}
& N^2(r) = - \mu J + \frac{r^2}{\ell^2} + \frac{\mu^2 \ell^4 M^2}{4 r^2}, \\
& \Omega_0(r) = \frac{\mu \ell^2 M}{2 r^2}, \\
& r_\pm = \frac{\ell}{2} \left[ \sqrt{\mu(J+M\ell)} \pm \sqrt{\mu(J-M\ell)} \right].
\end{align}
We now define a chiral black hole which is either a right-handed
rotating black hole of $0 \le M\ell \le J $ for the positive
Chern-Simons coefficient or a left-handed rotating black hole of $0
\le M \ell \le -J $ for the negative one. We assumed that the black
hole mass is positive, $M \ge 0$. So, the handedness of angular
momentum and angular velocity are associated with the signature of the
Chern-Simons coefficient.  As for the static black hole, it can be
achieved by $M=0$, which is actually massless.  The black hole can
surprisingly have a finite conserved angular momentum of the black
hole even in the massless case.  It means that the massless case is
not the vacuum state.  Note that the similar feature appears in the
BTZ solution in the CTMG in that the conserved angular momentum does
not vanish although the ADM mass is zero~\cite{kl,solodukhin}.  The
AdS spacetime can be also defined by $M=0,J=-1/\mu$ for each chiral
sector.  In connection with the vacuum state, it is regarded as an
empty space which can be realized by making the black hole disappear,
i.e., $M=0, J=0$.  So, the Hawking temperature and angular velocity
evaluated at the horizon $r_+$,
\begin{equation}
\label{tem}
T_H = \frac{r_+^2-r_-^2}{2\pi r_+ \ell^2}, \quad \Omega_H = \frac{r_-}{r_+ \ell},
\end{equation}
vanish for the vacuum state.

Next, we study the statistical entropy by using the brick wall method
of counting quantum modes near the horizon $r_+$~\cite{thooft}, then
the free energy of the quantum field is given by
\begin{equation}
F = -\frac{\zeta(3)}{\beta^3} \left( \frac{r_+^2 \ell^4}{(r_+^2-r_-^2)^2} \right) \frac{r_+}{\bar{h}},
\end{equation}
where $\bar{h}$ is the brick wall cutoff in terms of the proper length
and $\beta$ is an inverse temperature~\cite{kkps,hk}.  Then, fixing
the cutoff universally to $\bar{h}=3\zeta(3)/16\pi^3$ which is
independent of the mass and the angular momentum, the entropy is
obtained as
\begin{equation}
\label{S:bw}
S_\mathrm{bw} = \beta^2 \left. \frac{\partial F}{\partial \beta}
\right|_{\beta=\beta_H} 
= 4\pi r_+ ,
\end{equation}
which shows that this black hole entropy satisfies the area law.  Now
the internal energy $U$ and the angular momentum $J_m$ from this free
energy are easily calculated as
\begin{equation}
U = \frac{4}{3} \mu J - \frac{2r_-^2}{3\ell^2}, \quad J_m =  \mu \ell^2 M,
\end{equation}
respectively, then the thermodynamic first law of the quantum field
reads
\begin{equation}
\label{1st:bw}
\delta U = T_H \delta S_\mathrm{bw} + \Omega_H \delta J_m + (1/3\pi\ell^2) \left[ \delta A_+ - \delta A_- \right],
\end{equation}
where $A_\pm = \pi r_\pm^2$ are two-dimensional volumes of the inner
and outer horizons $r_\pm$.  Note that the thermodynamic first law can
be satisfied as long as we consider the pressure which is responsible
for the source that is nothing but the cosmological constant.  The
positive heat capacity can be also obtained as
\begin{equation}
\label{CJ:bw}
C_J^\mathrm{bw} = \left( \frac{\partial U}{\partial r_+} \right)_{J_m} \left( \frac{\partial T_H}{\partial r_+} \right)_{J_m}^{-1} = \frac{8\pi r_+ (2r_+^2-r_-^2)}{3 (r_+^2+3r_-^2)} > 0,
\end{equation}
where we have used a relation $(\partial r_- / \partial r_+)_{J_m} = -
r_-/r_+$.  As a result, we have obtained a thermodynamically stable
chiral black hole.

In fact, one can analyze the statistical entropy through the dual
conformal field theory (CFT) using the well-known Cardy
formula~\cite{cardy}.  The central charges and the conformal weights
can be read from the liming case of the CTMG in Ref.~\cite{kl},
$c_{L,R} = \mp (3\ell/2G_3) (\mu\ell)^{-1}$, and $h_{L,R} = (M\ell \mp
J)/2$, then the entropy becomes $S_\mathrm{CFT} = 2\pi \left[
  \sqrt{c_L h_L / 6} + \sqrt{c_R h_R / 6} \right] = 4 \pi r_+ / |\mu
\ell|$.  If the entropy expression is accepted, then one can easily
show the heat capacity is positive.  It is noteworthy that the entropy
can be written in the form of the outer horizon; however, one of the
two central charges should be negative so that the corresponding CFT
cannot possess the unitarity, and even more the conformal weight is
negative. In spite of this effort to derive the Bekenstein-Hawking's
area law, it is satisfactory up to the constant, $1 / \mu \ell$; more
worse, it fails to satisfy the thermodynamic first law, since it gives
$\delta M - \Omega_H \delta J = [2(r_+^2-r_-^2) / r_+ \mu \ell^3]
\delta r_- = T_H \delta S$, where the resulting entropy is $S= 4\pi
r_- / \mu\ell$. It is coincident rather with the Wald's
formula~\cite{wald:formula,tachikawa}, because the first law is
related to the conserved charges through the Noether theorem.  It can
be also shown that the entropy expressed by the Wald's formula does
not give the positive heat capacity.  So, the statistical entropy from
the dual CFT can be expressed by the outer horizon related to the
Bekenstein-Hawking's area law up to the constant and the heat capacity
is positive; however, it does not satisfies the thermodynamic first
law apart from the pathology of the negative central charge.

In conclusion, we have obtained a chiral black hole whose handedness
of the angular momentum depends on the signature of the Chern-Simons
coefficient. The third order derivative of the Cotton tensor gives a
dimensional constant which plays a role of the cosmological constant.
The conserved angular momentum and the mass of the chiral black hole
correspond to the mass and angular momentum of the BTZ black hole so
that the angular momentum can exist even though the angular velocity
is zero. On the other hand, the statistical entropy calculated by
using the brick wall method gives the Bekenstein-Hawking's area law,
thermodynamic first law, and the positive heat capacity while the CFT
side shows that the entropy is slightly different and the
thermodynamic first law fails.  It deserves to further study what the
genuine entropy is because the three expressions of 't Hooft (brick
wall), Cardy (CFT), and Wald (Noether theorem) are different.


\begin{acknowledgments}
We would like to thank S. Deser for comments and suggestions.
This work was supported by the Korea Science and Engineering Foundation 
(KOSEF) grant funded by the Korea government(MOST) 
(R01-2007-000-20062-0).
\end{acknowledgments}


\end{document}